\documentclass[conference]{IEEEtran}
\ifCLASSINFOpdf
\else
\fi

\usepackage{url}
\usepackage{amsmath}
\usepackage{subfigure}
\usepackage{multirow}
\usepackage{ctable}
\usepackage{booktabs}
\usepackage{hyperref}
\usepackage{setspace}

\newcommand{\squishlist}{
\begin{list}
	{$\bullet$} { \setlength{
	\itemsep}{0pt} \setlength{\parsep}{3pt} \setlength{\topsep}{3pt} \setlength{
	\partopsep}{0pt} \setlength{\leftmargin}{1.5em} \setlength{\labelwidth}{1em} \setlength{\labelsep}{0.5em} } }
	
	\newcommand{\squishlisttwo}{
	\begin{list}
		{$\bullet$} { \setlength{
		\itemsep}{0pt} \setlength{\parsep}{0pt} \setlength{\topsep}{0pt} \setlength{
		\partopsep}{0pt} \setlength{\leftmargin}{2em} \setlength{\labelwidth}{1.5em} \setlength{\labelsep}{0.5em} } }
		
		\newcommand{\squishend}{
	\end{list}
	}

\makeatletter
\def\@IEEEpubidpullup{8\baselineskip}
\makeatother
\begin{document}
%

\IEEEoverridecommandlockouts
\IEEEpubid{
	\parbox{\columnwidth}{\vspace{-4\baselineskip}Permission to make digital or hard copies of all or part of
		this work for personal or classroom use is granted without fee provided that copies are not made or
		distributed for profit or commercial advantage and that copies bear this notice and the full citation on the
		first page. Copyrights for components of this work owned by others than ACM must be honored.
		Abstracting with credit is permitted. To copy otherwise, or republish, to post on servers or to redistribute
		to lists, requires prior specific permission and/or a fee. Request permissions from
		\href{mailto:permissions@acm.org}{permissions@acm.org}.\hfill\vspace{-0.8\baselineskip}\\
		\begin{spacing}{1.2}
			\small\textit{ASONAM '17}, July 31 - August 03, 2017, Sydney, Australia \\
			\copyright\space2017 Association for Computing Machinery. \\
			ACM ISBN 978-1-4503-4993-2/17/07?/\$15.00 \\
			\url{http://dx.doi.org/10.1145/3110025.3110069}
		\end{spacing}
		\hfill}
	\hspace{0.9\columnsep}\makebox[\columnwidth]{\hfill}}
\IEEEpubidadjcol

\title{Identifying On-time Reward Delivery Projects with Estimating Delivery Duration on Kickstarter}

\author{\IEEEauthorblockN{Thanh Tran$^\dagger$, Kyumin Lee$^*$, Nguyen Vo$^\dagger$, Hongkyu Choi$^\dagger$}
\IEEEauthorblockA{$^\dagger$Department of Computer Science, Utah State University, Logan, UT, USA \\
	$^*$Department of Computer Science, Worcester Polytechnic Institute, Worcester, MA, USA \\
 \{thanh.tran,nguyenvo,hongkyu.choi\}@aggiemail.usu.edu, kmlee@wpi.edu}
}


%


\maketitle

\begin{abstract}
In Crowdfunding platforms, people turn their prototype ideas into real products by raising money from the crowd, or invest in someone else's projects. In reward-based crowdfunding platforms such as Kickstarter and Indiegogo, selecting accurate reward delivery duration becomes crucial for creators, backers, and platform providers to keep the trust between the creators and the backers, and the trust between the platform providers and users. According to Kickstarter, 35\% backers did not receive rewards on time. Unfortunately, little is known about on-time and late reward delivery projects, and there is no prior work to estimate reward delivery duration. To fill the gap, in this paper, we (i) extract novel features that reveal latent difficulty levels of project rewards; (ii) build predictive models to identify whether a creator will deliver all rewards in a project on time or not; and (iii) build a regression model to estimate accurate reward delivery duration (i.e., how long it will take to produce and deliver all the rewards). Experimental results show that our models achieve good performance -- 82.5\% accuracy, 78.1 RMSE, and 0.108 NRMSE at the first 5\% of the longest reward delivery duration.
\end{abstract}


%
\IEEEpeerreviewmaketitle

\section{Introduction}

Crowdfunding platforms have successfully connected millions of individual investors to creators, and helped creators to bring their ideas into the reality. In recent years, a market size of crowdfunding platforms has increased exponentially, reaching tens of billions of dollars. Among various types of crowdfunding platforms, reward-based crowdfunding platforms have become popular, especially, Kickstarter has become the most popular crowdfunding platform. According to Kickstarter\footnote{https://www.kickstarter.com/help/stats}, more than 2.5 billion dollars were pledged by approximately 12 million backers to more than 110k projects.

\begin{figure*}
	\centerline{
		\includegraphics[width=0.86\linewidth]{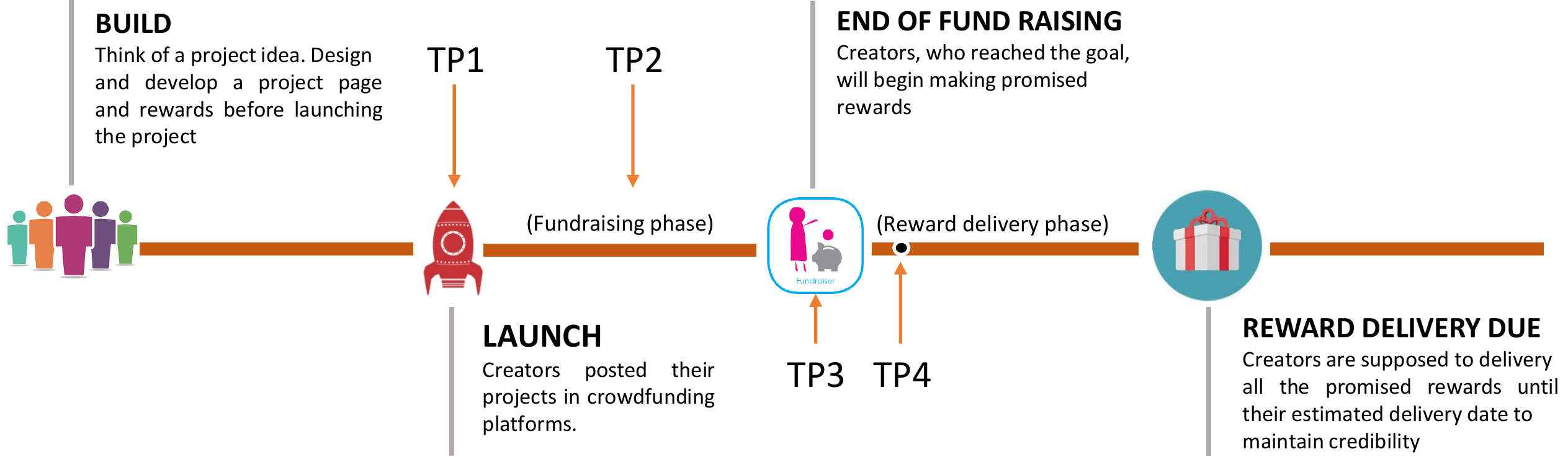}
	}
	\vspace{-5pt}
	\caption{Project Timeline. In this study, we build our models at four time points (TP1, TP2, TP3 and TP4).}
	\label{fig:campaignTimeline}
	\vspace{-10pt}
\end{figure*}

As shown in Figure~\ref{fig:campaignTimeline}, a project in reward-based crowdfunding platforms has two phases: (1) \emph{the fundraising phase} -- when a creator raises money by promoting the project after launching it; and (2) \emph{the reward delivery phase} -- when the creator makes and ships products as the rewards if the project was successful in terms of pledged money $\geq$ goal.

In the literature, researchers mostly focused on \emph{the fundraising phase}, by analyzing dynamics of crowdfunding platforms \cite{kuppuswamy2015crowdfunding}, understanding why people created projects or backed other projects \cite{gerber2012crowdfunding}, studying how to make a project successful \cite{hui2014understanding,lu2014identifying,lu2014inferring,tran2016succeed,xu2014show}, predicting project success \cite{chung2015long,etter2013launch,greenberg2013crowdfunding,li2016project}, and recommending creators to backers or vice versa \cite{an2014recommending,rakesh2016probabilistic}. However, researchers rarely paid attention to \emph{the reward delivery phase}.

According to Kickstarter\footnote{https://www.kickstarter.com/fulfillment}, 35\% backers did not receive rewards on time. If creators send rewards to backers on time, backers will be likely to invest in their upcoming projects \cite{althoff2015donor}. Although the time already passed the fundraising phase, if creators announce production and delivery delay with a new estimated date as soon as possible, some backers will still wait for receiving the rewards without losing much trust and without much surprise. Some backers may request a refund to creators without waiting until the estimated date (e.g., 1 year).

While on-time reward delivery becomes crucial for retaining backers in the creators' future projects, it is difficult for creators to estimate an accurate delivery date because of various reasons. First, 90\% creators created a project the first time, so they don't have much experience in accurately estimating delivery date \cite{tran2016succeed}. Second, some creators choose a delivery date with their hunch without understanding the reward's difficulty level. Third, there may be other uncertainties like factory issues and unexpected problems in their prototypes, requiring more time.

Unfortunately, little is known about what factors influence to on-time or late reward delivery projects, and there is no prior work to estimate reward delivery duration. To fill the gap, in this paper, we focus on answering following research questions: Can we build a predictive model which can predict whether a project will be an on-time delivery project or not? Can we build a model which can estimate delivery duration accurately? Completing these two tasks would be a big challenge with only using observable online data available in a crowdfunding platform.

To answer the research questions, first we defined four time points \emph{TP1}, \emph{TP2}, \emph{TP3} and \emph{TP4}, which are when a project is launched, the middle of the fundraising phase, the end of the fundraising phase, and the first 5\% of the estimated longest reward delivery duration, respectively as shown in Figure~\ref{fig:campaignTimeline}. An ideal model is supposed to predict on-time or delay at \emph{TP1} and \emph{TP2} well so that the investors can decide whether they are going to back the project or not. However, in practice, it would be very difficult because of many uncertainties and limited data. Therefore, building models at \emph{TP3} and \emph{TP4} are important and valuable as long as it can achieve high accuracy because creators can announce their delay or re-estimate reward delivery date at \emph{TP3} and \emph{TP4}. In addition, backers can request issuing refund at \emph{TP3} and \emph{TP4}. According to the refund policy of Kickstarter, it is possible for creators to refund anytime, and backers can request a refund during \emph{the reward delivery phase}.

In this paper, we make the following contributions:
\begin{itemize}
    \item First, we proposed a clustering approach to group rewards by their latent difficulty levels, generating new features. We also extracted distinguishing characteristics of projects, creators and backers in on-time and late reward delivery projects.
    \item Second, based on the analysis and extracted features, we developed predictive models to classify on-time and late reward delivery projects at \emph{TP1}, \emph{TP2}, \emph{TP3} and \emph{TP4}.
    \item Finally, we proposed and developed a predictive model to estimate delivery duration accurately at \emph{TP1}, \emph{TP2}, \emph{TP3} and \emph{TP4}. To our knowledge, this is the first work to study how to predict a project delivery status (on-time or late) and estimate delivery duration.
\end{itemize}

\vspace{-1mm}
\section{Related Work}

In this section, we summarize prior research works related to the analysis of projects and users, project success prediction and recommendation on crowdfunding platforms.

Researchers analyzed crowdfunding platforms \cite{Belleflamme:2012,Gerber:2013}. For example, Kuppuswamy et al. \cite{kuppuswamy2015crowdfunding} showed the dynamics of Kickstarter donors. Mollick et al. \cite{mollick:2014} studied the dynamics of crowdfunding and revealed that personal networks and underlying project quality were related to the crowdfunding success. Gerber et al. \cite{gerber2012crowdfunding} analyzed why people created and/or backed projects in crowdfunding platforms. Xu et al. \cite{tran2016succeed,xu2014show} showed various factors to make a project successful in terms of the fundraising. Joenssen and M{\"u}llerleile \cite{joenssen2016limitless} analyzed 42k Indiegogo projects, and discovered that scarcity management was problematic and reduced the chances of projects to successfully achieve the fundraising goal. Researchers \cite{hui2014understanding,lu2014identifying,lu2014inferring} studied the impact of social media and social communities in raising fund. Joenssen et al. \cite{joenssen2014link} showed that timing and communication were two key factors to make projects successful.

Etter et al. \cite{etter2013launch} examined 16k Kickstarter projects and proposed a model based on pledged money features, and project and backer graph features to predict the success of the projects. Greenberg et al. \cite{greenberg2013crowdfunding} extracted 13 features from each of 13,000 Kickstarter projects and developed classifiers to predict project success. In \cite{chung2015long,tran2016succeed}, the authors used different feature traits to predict the success of projects. Li et al. \cite{li2016project} analyzed 18K Kickstarter projects and built logistic and log-logistic based models to predict the chance of successfully achieving goal. Solomon et al. \cite{solomon2015don} discovered that early donation played an important role in making the project successful. Mitra et al. \cite{Mitra:2014} proposed text features of project pages for the project success prediction.

Other researchers studied building recommender systems for creators and backers/investors. An et al. \cite{an2014recommending} analyzed backers' pledging behavior, and built a SVM classifier to suggest potential backers to creators. In \cite{rakesh2015project}, the authors used temporal, personal, geo-location and network traits to recommend a set of potential backers to projects. Rakesh et al. \cite{rakesh2016probabilistic} examined a project status, personal preference of individual investors and preference of investor groups. Then, they proposed a probabilistic recommendation model to recommend projects to a group of investors.

Recently, Kim et al. \cite{kimunderstanding} interviewed crowdfunding participants and found various factors which influenced backers' trust. They also conducted analysis for 4,089 delayed projects to understand how 8 factors were related to delayed duration. However, they did not study which project will pass estimated delivery date, nor how long a reward delivery including production will take.


While most of the previous research works focused on \emph{the fundraising phase}, we focus on the next phase -- \emph{reward delivery phase}. Our work is also different from \cite{kimunderstanding}. In particular, we analyze characteristics of on-time and late delivery projects, and build a predictive model to predict whether rewards in a project will be delivered on-time or not. In addition, we build a regression model to recommend accurate delivery duration to creators.
\vspace{-1mm}
\section{Dataset}
\label{sec:dataset}

This section presents our dataset with two types of ground-truth: (1) on-time or late reward delivery project and (2) actual delivery duration.

\noindent\textbf{Ground truth collection:} We define an on-time reward delivery project and a late reward delivery project as follows:
\squishlist
\item On-time reward delivery project: If all rewards in a project were shipped by the longest estimated delivery date (LEDD), it would be called an on-time reward delivery project. Note that a project creator decides each reward's estimated delivery date when she creates her project page. 

\item Late reward delivery project: If a creator did not ship at least one of rewards by the LEDD, the project would be called a late reward delivery project.
\squishend

We collected all project pages (i.e., 168,851 project pages) from Kickstarter which were created between 2009 and September 2014. Among the 168,851 projects, we extracted successful projects, each of which had a project goal equal to or greater than \$100. 29,499 successful projects satisfied the condition. In addition, we collected updates and comments associated with the successful projects.

Labeling each project for delivery status and duration requires reading all the updates and comments. Instead of labeling all the successful projects, we sampled 10\% of the 29,499 successful projects with keeping the same project distribution over project categories, year and goal. Then, three labelers independently labeled the 2,949 sampled projects based on the following guideline:
\squishlist
\item If a labeler could identify that all rewards in a project were shipped by LEDD (based on updates and comments), and there was no complaint regarding not receiving the rewards, she would label the project as an on-time reward delivery project.

\item If there was at least one update from a creator after LEDD regarding delayed shipping or a comment with a new delivery date beyond LEDD, a labeler labeled the project as a late reward delivery project.
\squishend

\begin{figure}[t]
\centering
\includegraphics[width=0.8\linewidth]{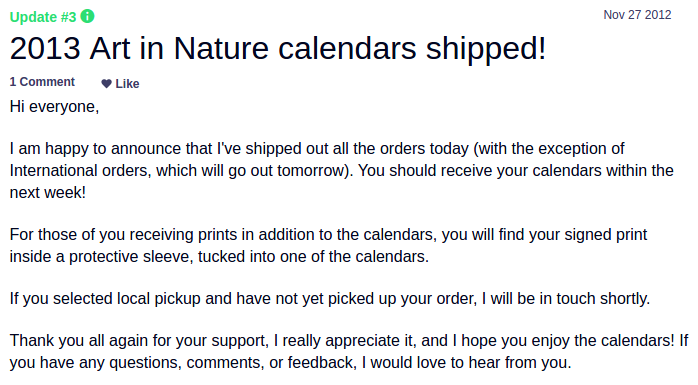}
\vspace{-10pt}
\caption{An update containing shipping information.}
\label{fig:infered-delivery}
\vspace{-10pt}
\end{figure}

We excluded projects if labelers were not able to verify whether a project is an on-time reward delivery project or not based on the labeling guideline. Finally, 2,198 projects were labeled by them, and consisted of 1,003 on-time and 1,195 late reward delivery projects.

Next, we were interested in collecting true/actual delivery duration as the ground truth (i.e., how long it took to deliver all the rewards since the end of the fundraising phase). Out of 2,198 projects, the creators of 1,598 projects posted updates with information when they shipped all the rewards, as shown in Figure~\ref{fig:infered-delivery}. Based on the information, true/actual delivery duration of the 1,598 projects was calculated.

\begin{figure} [t]
	\centering
	\includegraphics[width=0.76\linewidth]{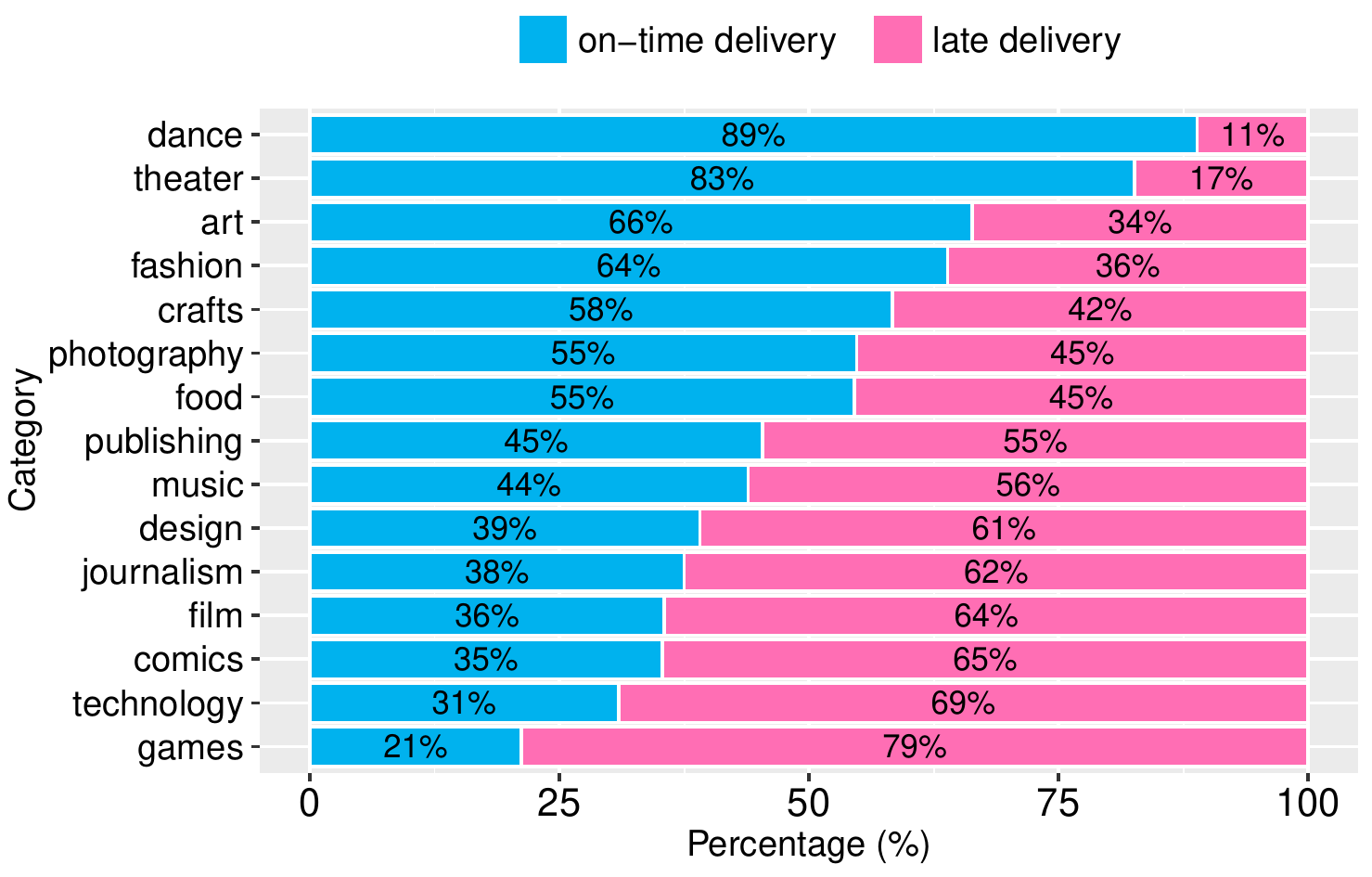}
	\vspace{-10pt}
	\caption{Category distributions of on-time and late reward delivery projects.}
	\label{fig:OntimePassedProjectsDistribution}
	\vspace{-10pt}
\end{figure}

\noindent\textbf{Categorical distribution:} Figure~\ref{fig:OntimePassedProjectsDistribution} shows the categorical distributions of on-time and late reward delivery projects. There was a higher probability of on-time reward delivery in dance and theater-related projects because the rewards in those categories were often live performance, show cases or dancing tutor classes, and were served at once for all backers. In contrast, the rewards in other categories like games, technology, film were real products (e.g., a game, book, movie), requiring more time to produce and deliver to backers.
\vspace{-1mm}
\section{Clustering Rewards by the Latent difficulty Level}
\label{sec:semanticClustering}

In this section, we propose a novel approach to measure a reward's difficulty level and a project's overall difficulty level toward extracting features, which will be a part of our final feature set for building models in the following sections. Our hypothesis is that true delivery duration for a reward depends on its difficulty level, and reward description may reveal the difficulty level. It makes sense that developing a game as a reward requires more time and effort than producing a t-shirt. In this section, we study how to measure the difficulty level of each reward and represent how hard a project is in terms of producing and delivering its rewards.

\smallskip
\noindent\textbf{Clustering approach to get new features:} We group rewards into the same cluster if their descriptions contain semantically similar meaning. Intuitively, if two reward descriptions are semantically similar, they may have similar difficulty level and thus require a similar amount of time to produce and deliver.

Our approach consists of six steps as follows:

\begin{itemize}
\item Step 1: 1,273,617 rewards were extracted from 149,189 Kickstarter projects, and their reward descriptions were preprocessed by removing stop words and punctuation.
\item Step 2: Using 1,273,617 reward descriptions, we built Glove model \cite{pennington2014glove}, in which each word is represented by a vector. In our implementation, we set up vector size=50, maximum number of iterations=20, window size=15, and vocabulary minimum count=5.
\item Step 3: From Step 2, all words in the 1,273,617 reward descriptions were represented by Glove vectors. We grouped the words into \emph{K1} clusters by running k-means clustering algorithm. To choose the optimal \emph{K1}, we varied \emph{K1} from 1 to 100 and selected the value that minimized BIC value as follows:
\begin{equation}
\label{equa:BICformula}
\vspace{-3pt}
BIC = \sum_{k=1}^{K1}\sum_{i \in words_k}^{}dist(v_i,c_k) + log(n)*m*K1
\vspace{-3pt}
\end{equation}
where $v_i, c_k$ is the representative vector of the word $i$ in cluster $k$ and the center of cluster $k$, respectively, $dist(v_i,c_k)$ is the Euclidean distance of two vectors $v_i$ and $c_k$, $n$ is the number of rewards (e.g. $n = 1,273,617$), $m$ is the number of dimensions of the word vector, and $K1$ is the number of clusters. Finally, the optimal K1 was 67. So at step 3, we clustered words into 67 groups.

\item Step 4: From the sampled and labeled 2,198 projects, we extracted 19,266 reward descriptions.
\item Step 5: We represented each of 19,266 reward descriptions to a vector with 67 dimensions, each of which is mapped with a word cluster in Step 3. In particular, we counted how many words in each cluster occurred in the reward description, and used the count as a value of the dimension mapped to the cluster.

\item Step 6: Each of 19,266 rewards/reward descriptions was represented by a vector in 67 dimensions. Then, we clustered the rewards into 14 groups. Like Step 3, we did the same process finding the optimal number of clusters. We call the 14 groups as 14 \emph{semantic reward clusters}.
\end{itemize}

By doing the six steps, we got 14 semantic reward clusters. 
Then, we generated 14 feature values for each of 2,198 projects as follows: given a project $p_k$, rewards in $p_k$ and corresponding number of backers to each reward, we summed up the number of backers of each reward that belongs to $i$th semantic cluster $c_i$, and used it as a feature value. We considered both a difficulty level of each reward and the corresponding number of backers because more backers mean the creator has to produce more number of outcomes. Finally, the project $p_k$ was represented by a vector in 14 dimensions.

\smallskip
\noindent\textbf{Quality of the clusters:} To prove that each semantic reward cluster has a distinguishing difficulty level, first we identified each project $p_k$'s major semantic cluster $M(p_k)$, a cluster (i.e., one of the 14 clusters) to which the large number of rewards in $p_k$ belongs. Then, given a set of on-time delivery projects $A$, we defined an indicator function $\textbf{1}_A$ of the project $p_k$ as follows:
\[
		\textbf{1}_A(p_k) = \begin{cases}
								1 	&		\text{if $p_k \in A$}  \\
								0, 	&		\text{if $p_k \notin A$}
						\end{cases}
\]

Then, given a probability of each semantic reward cluster $P(c_i)$ ($=\frac{|projects \; having \; c_i \; as \; the \; major \; semantic \; cluster|}{total \; number \; of \; projects}$), we calculated the conditional probability of the project $p_k$ to be an on-time delivery project using Bayes theorem as follows:
\[
\resizebox{0.45\textwidth}{!}{$
P(\textbf{1}_A(p_k) = 1 | c_i = M(p_k)) = \frac{P(c_i=M(p_k)|\textbf{1}_A(p_k) = 1) * P(\textbf{1}_A(p_k) = 1)}{P(c_i = M(p_k))}
$}
\]
where $P(c_i = M(p_k))$ is P($p_k$'s major semantic cluster).

\begin{figure} 
	\centering
	\includegraphics[width=0.36\textwidth]{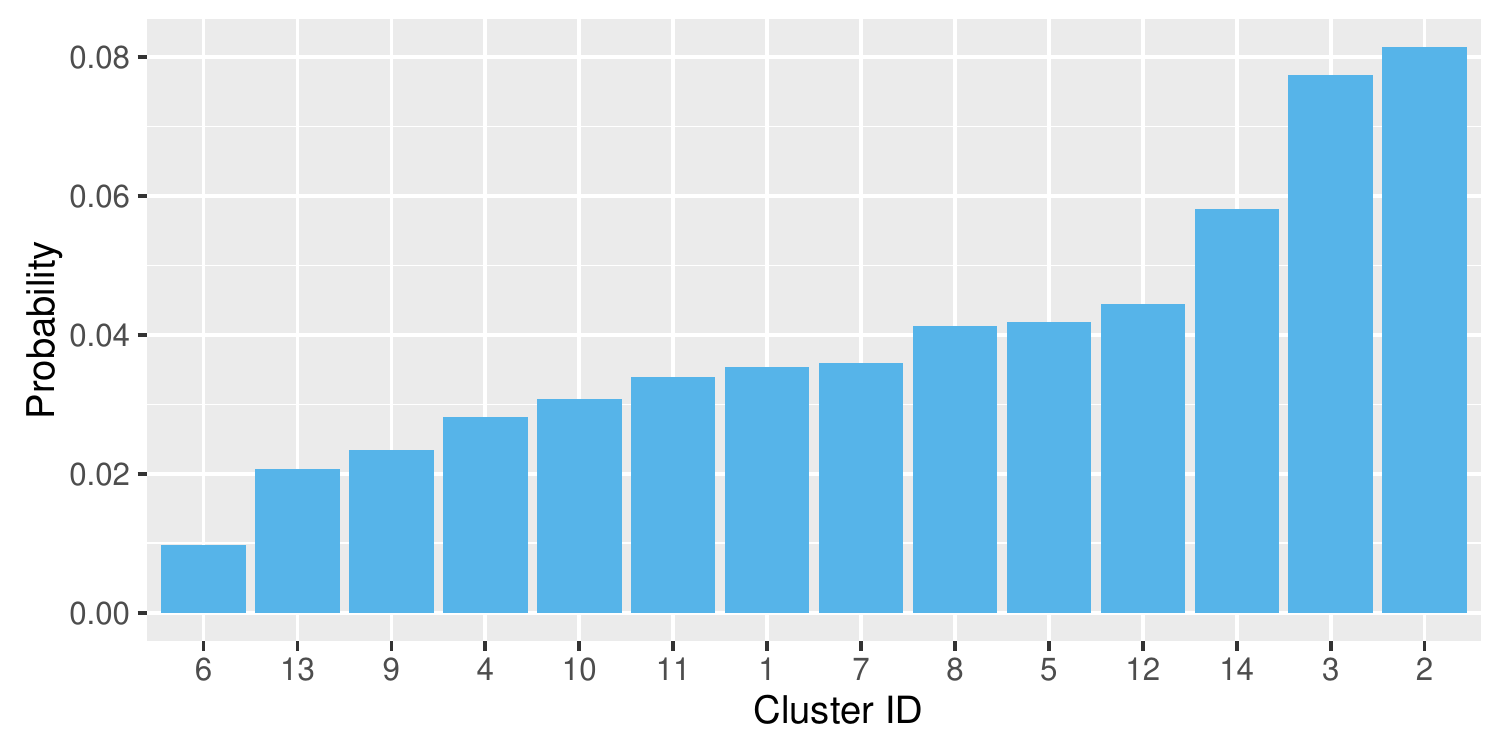}
	\vspace{-10pt}
	\caption{14 semantic reward clusters with their on-time delivery conditional probabilities in the descending order of difficulty levels.}
	\label{fig:ProbFig}
	\vspace{-10pt}
\end{figure}

\begin{figure} 
\centering
	\subfigure[Easy level] 
	{
		\label{fig:WordCloud-SemanticCluster-easy}
		\includegraphics[width=0.12\textwidth]{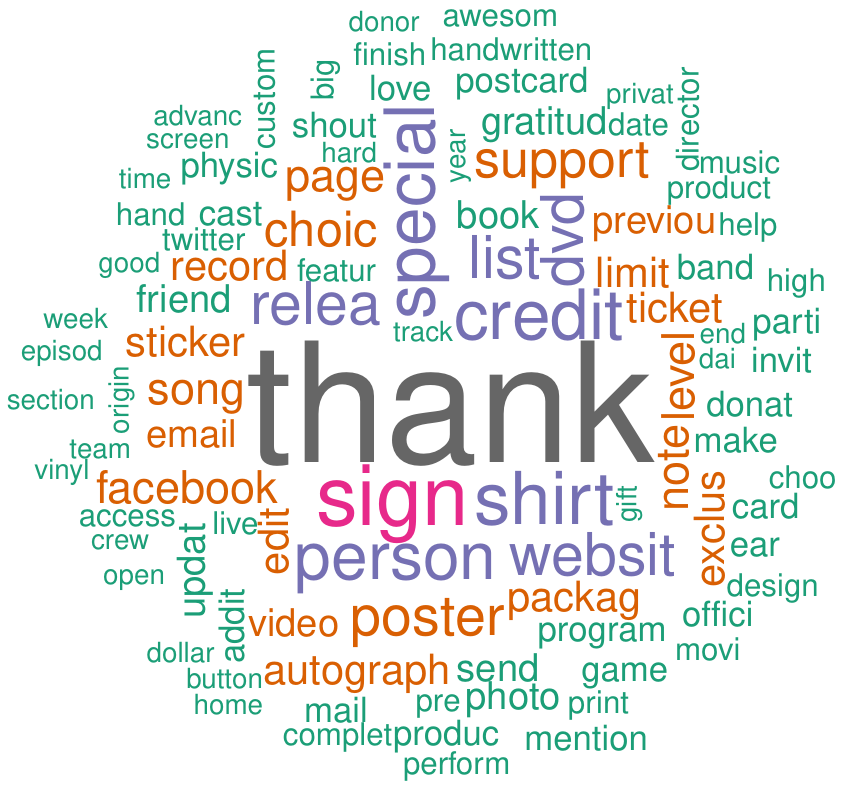}
	}
	\subfigure[Medium level] 
	{
		\label{fig:WordCloud-SemanticCluster-medium}
		\includegraphics[width=0.12\textwidth]{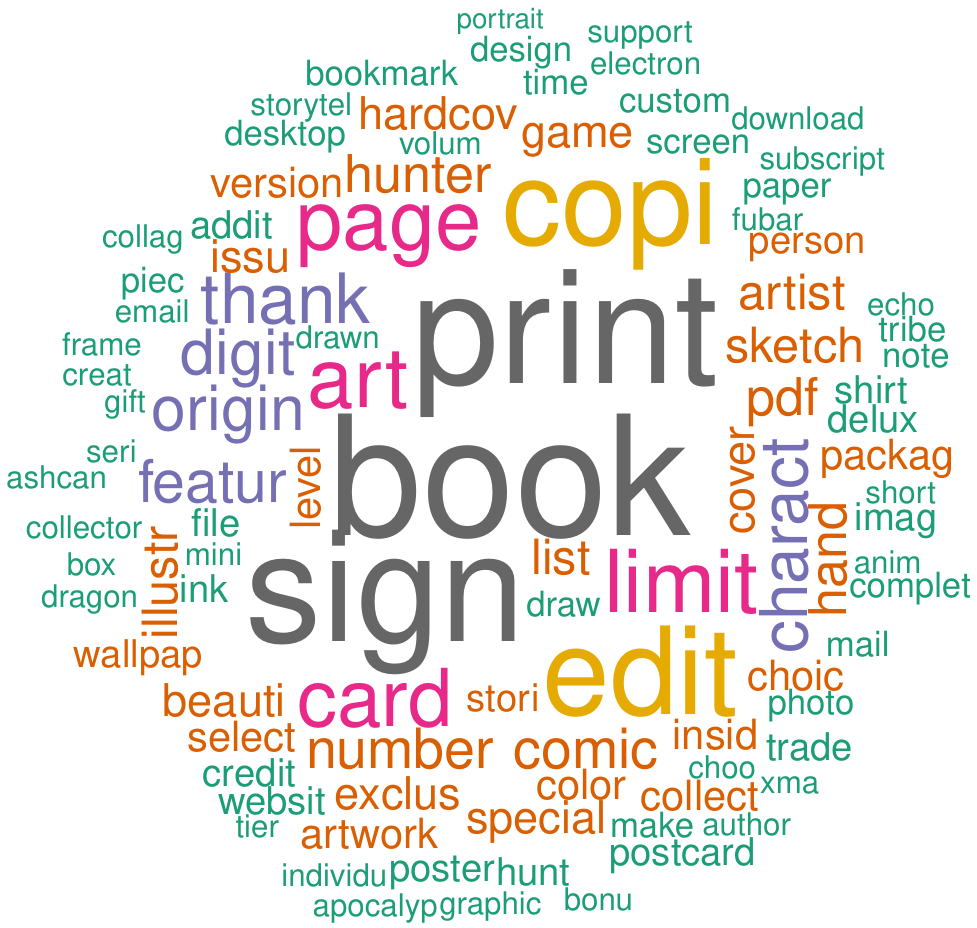}
	}
	\subfigure[Hard level] 
	{
		\label{fig:WordCloud-SemanticCluster-hard}
		\includegraphics[width=0.12\textwidth]{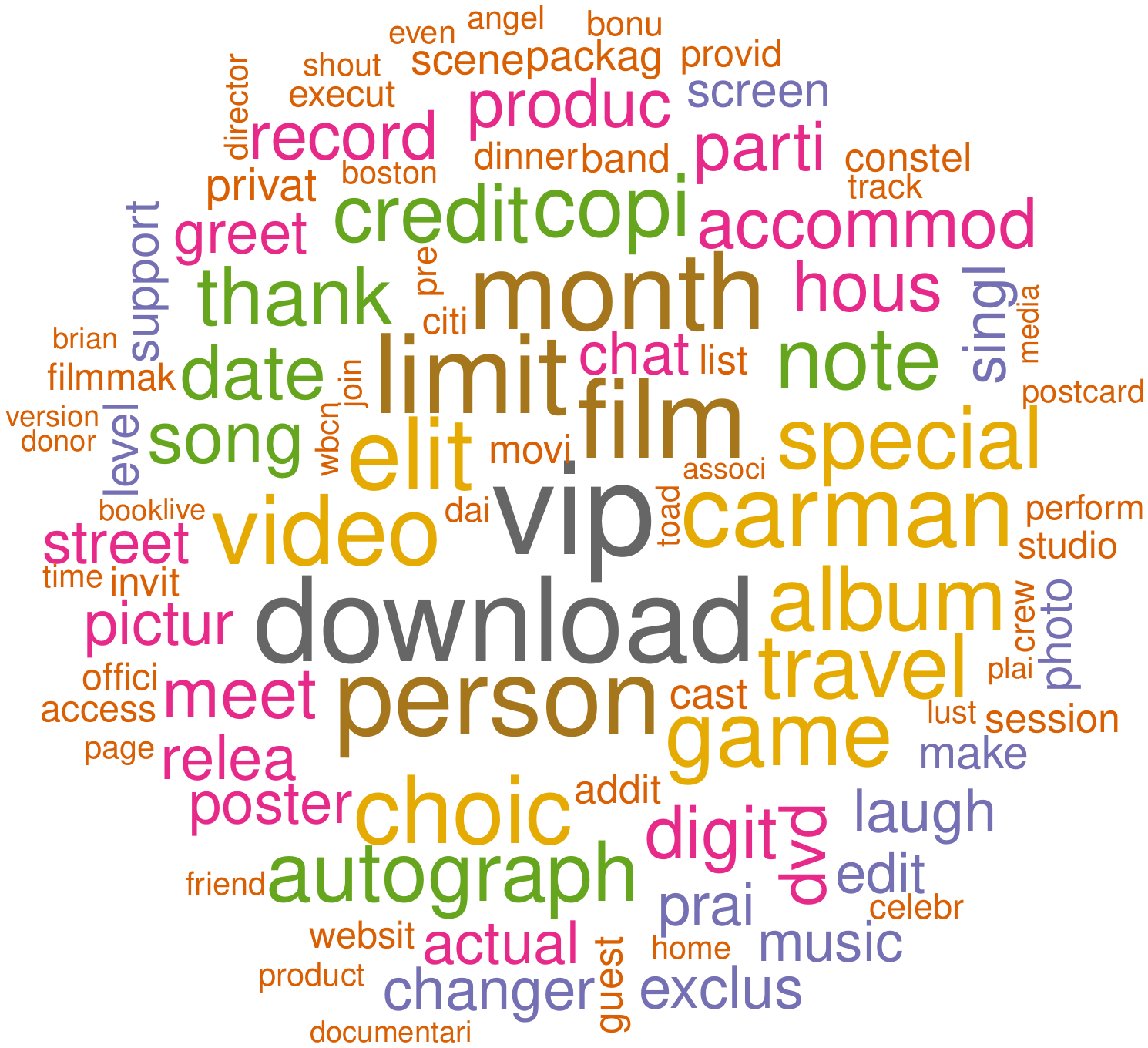}
	}
	\vspace{-5pt}
	\caption{Word clouds of three clusters: cluster 2 (easy level), cluster 1 (medium level) and cluster 6 (hard level).}
	\label{fig:WordCloud-SemanticClusters}
	\vspace{-10pt}
\end{figure}

Figure \ref{fig:ProbFig} presents the conditional probability of $P(\textbf{1}_A(p_k) = 1 | c_i = M(p_k))$ in each semantic reward cluster by descending order of difficulty level. Each cluster had a different probability. Semantic reward cluster \emph{6} had the lowest probability, indicating that rewards in this group had a higher difficulty level than other groups (e.g. hard level). In contrast, semantic reward cluster \emph{2} had the highest probability, showing that the rewards in this cluster were easier in terms of producing and shipping (e.g. easy level). Semantic reward cluster \emph{1} had a middle probability (e.g. medium level). We next plotted the word clouds of those three clusters to understand what kind of rewards were included in the three clusters. In Figure \ref{fig:WordCloud-SemanticClusters}, we observed that the rewards in easy level (cluster 2) mostly contained keywords like ``thank, credit, shirt, sign, websit'' which can be delivered quickly. Rewards in the medium level (cluster 1) contained keywords like ``print, book, sign, copi'', related to publishing category. Rewards in the hard level (cluster 6) contained keywords like ``film, vip, video, game'', related to the film and game category. As shown in Figure~\ref{fig:OntimePassedProjectsDistribution}, it makes sense that film, games were those categories with the highest late delivery rate, whereas publishing category had a lower late delivery rate. In the following section, we show that adding semantic cluster features improved our prediction rate.

\begin{table} [t]
	\centering
	\vspace{-5pt}
	\caption{Features that were newly extracted at each time or phase.}
    \small
	\label{table:features}
	\resizebox{0.39\textwidth}{!}{
	\begin{tabular}{|p{3.2in}|}
		\hline
		\textbf{At the launching time} \\ \hline
		\emph{Project based features}: $|$images$|$, $|$faqs$|$, goal, project category, $|$rewards$|$, $|$reward sentences$|$, $|$bio description sentence$|$, fund raising duration, the longest reward delivery duration, SMOG score of project, reward and bio description, and semantic reward clustering features.\\
		\hline \hline

		\textbf{During the fundraising phase} \\ \hline
		\emph{Project based features}: $|$backers$|$, $|$project's comments$|$, $|$project's updates$|$.\\
		\emph{Creator's activeness features}: $|$creator's comments$|$, $|$creator's updates$|$.\\
		\emph{Temporal features}: $|$comments$|$ in each of 20 time slots.\\
        \hline \hline

		\textbf{At the first 5\% of the longest reward delivery duration} \\ \hline
		\emph{Creator's activeness features}: $|$creator's comments$|$, $|$creator's updates$|$, average update time interval, and average response time between a backer's question and a reply from the creator.\\
		\emph{Backer's activeness features} : $|$backers' comments$|$, $|$backers who posted comments$|$, $|$backers' questions$|$.\\
		\emph{Linguistic features:} LIWC feature extracted from updates (12 scores), LIWC feature extracted from comments (5 scores). \\
		\hline
	\end{tabular}
	}
    \vspace{-10pt}
\end{table}

\begin{table*} [t]
	\centering
	\vspace{-5pt}
	\caption{Top 10 features at TP1, TP2, TP3 and TP4}
\small
	\label{table:featuresRanking}
	\resizebox{0.9\textwidth}{!}{
	\begin{tabular}{lc|lc|lc|lc}
		\toprule
		\multicolumn{2}{c}{Launching time (TP1)} & \multicolumn{2}{c}{Middle of fundraising (TP2)} & \multicolumn{2}{c}{End of fundraising (TP3)} & \multicolumn{2}{c}{First 5\% of reward delivery duration (TP4)}  \\
		\midrule
		Features      								& z-score & Features									& z-score 	&
		Features									& z-score & Features 									& z-score   \\
		\midrule
		project category 							& 24.11	&	project category							& 22.93 	&
		$|$updates$|$								& 22.47	&	$|$creator's updates$|$ (at TP4)			& 35.73 	\\
														
		goal										& 18.63	&	goal										& 15.24		&
		project category							& 17.47 &	$|$creator's comments$|$ (at TP4)			& 18.19 	\\
		
		longest reward delivery duration			& 13.40	&	longest reward delivery duration			& 10.99 	&
		goal										& 12.76	&	project category							& 14.80 	\\
		
		smog score of project's description			& 9.71	&	$|$temporal Comment at slot $1^{st}$$|$		& 8.52 		&
		$|$creator's comments$|$					& 11.70	&	longest reward delivery duration			& 13.60 	\\
						
		semantic reward cluster $9^{th}$			& 7.26	&	$|$temporal Comment at slot $8^{th}$$|$		& 8.38 		&
		longest reward delivery duration			& 10.23	&	$|$backer's comments$|$ (at TP4)			& 13.16	 	\\
		
		semantic reward cluster $2^{nd}$			& 6.67	&	$|$temporal Comment at slot $9^{th}$$|$		& 8.29 		&
		$|$temporal Comment at slot $18^{th}$$|$	& 7.65	&	$|$project's comments$|$					& 12.69 	\\
		
		$|$images$|$								& 5.90	&	$|$temporal Comment at slot $2^{nd}$$|$				& 7.94 		&
		$|$temporal Comment at slot $19^{th}$$|$	& 7.43	&	average update time interval				& 12.07  	\\
		
 		$|$Faqs$|$									& 4.97	&	$|$temporal Comment at slot $7^{th}$$|$		& 7.74 		&
 		$|$temporal Comment at slot $17^{th}$$|$	& 7.08	&	goal 										& 10.76	 	\\
 		
	 	semantic reward cluster $12^{th}$			& 4.89	& 	$|$temporal Comment at slot $4^{th}$$|$		& 7.73 		&
	 	$|$temporal Comment at slot $14^{th}$$|$	& 6.92	&	$|$creator's comment$|$ (at TP3)			& 8.16	 	\\
	 	
	 	semantic reward cluster $7^{th}$			& 4.45	& 	smog score of project's description			& 7.67 		&
	 	$|$temporal Comment at slot $16^{th}$$|$	& 6.88	&	$|$temporal comment at slot $18^{th}$$|$	& 6.53	 	\\

		\bottomrule
		\hline
	\end{tabular}
	}
	\vspace{-10pt}
\end{table*}

\vspace{-1mm}
\section{Feature sets}
Toward identifying on-time and late reward delivery projects as well as estimating reward delivery duration for the projects, we extracted the following features and used in the following sections:
\begin{itemize}
\item{Project-based features:} We extracted 16 project-related features: $|$images$|$, $|$faqs$|$, goal, project category, $|$rewards$|$, $|$reward sentences$|$, $|$bio description sentence$|$, fund raising duration, the longest reward delivery duration, SMOG score \cite{smogscore:Misc} of project description, SMOG score of reward description, SMOG score of bio description,  $|$backers$|$, $|$project's comments$|$, $|$project's updates$|$, and semantic reward clustering feature.
\item{Creator activeness features:} We extracted 4 features related to the creator's activeness: $|$creator's comments$|$, $|$creator's updates$|$, average update time interval, and average response time between a backer's question and a reply from the creator.
\item{Backer's activeness features:} Backers can post comments and ask for the project progress. We extracted 3 features related to the backer's activeness: $|$backers' comments$|$, $|$backers who posted comments$|$, $|$backers' questions$|$.
\item{Temporal features:} We converted each project’s fundraising duration into 20 states (time slots), since projects have various fundraising periods (e.g., 30 days, 60 days). In each state/time slot, we measured the number of comments posted by creators and backers.
\item{Linguistic usage of creators and backers:} We used Linguistic Inquire and Word Count (LIWC) dictionary \cite{pennebaker2007linguistic} to discover distinguished linguistic usage patterns of creators (through their updates). To compute the linguistic usage score of creators in on-time reward delivery projects and creators in late reward delivery projects over 64 LIWC categories, we performed the same process which was mentioned in \cite{lee2014will,lee2013crowdturfers}. We applied two-sample t-test and assigned $\alpha$ as 0.00078 (=0.05/64) to select only the LIWC categories in which we observed a significant difference between two distributions. Finally, we found 12 LIWC categories in which creators in on-time and late reward delivery projects had a significant linguistic-usage difference. Via the same process, we measured the different linguistic usage of backers by their comments. We found 5 LIWC categories in which backers in on-time and late reward delivery projects had a significant linguistic-usage difference.
\end{itemize}

\vspace{-1mm}
\section{Identifying On-time and Late Reward Delivery Projects}
\label{sec:experiment1}

In this section, we build predictive models to classify whether a project is an on-time delivery project or not and evaluate their performance against baselines.


\vspace{-1mm}
\subsection{Experimental Setting}
As shown in Figure~\ref{fig:campaignTimeline}, we conducted experiments at the four time points, depending on what information available at each time point: (1) \textbf{TP1}: when a project is launched; (2) \textbf{TP2}: in the middle of the fundraising phase; (3) \textbf{TP3}: in the end of the fundraising phase; and (4) \textbf{TP4}; at the first 5\% of the longest reward delivery duration (to see whether building classifiers in 5\% delivery duration improve the classification performance compared with TP1, TP2 and TP3). At each time point, we extracted available features and conducted 10-fold cross-validation. Table~\ref{table:features} presents our proposed features. We added previously available features to following time points.

At each time point, we did feature selection by removing linearly related features and unimportant features. Particularly, to remove linearly related features, we measured the \emph{variance inflation factor} (VIF) of each feature. VIF value of a feature $i$ is computed as follows:
\[
\vspace{-3pt}
	VIF_i = \frac{1}{1 - R^2_i}
\vspace{-3pt}
\]
where $R^2_i$ is the coefficient of multiple determination obtained by doing regression of the feature $i$ as response to the remaining features. A VIF value is in a range of $[0, \infty)$. If a feature $i$'s VIF value is 1, it means there is no correlation between the feature and the other features. But, if its VIF value is equal to 10 or greater, it means there exists multicollinearity. Removing the multicollinearity follows three steps: \emph{Step 1}: VIF values of all \emph{n} features are computed; \emph{Step 2}: If some features have VIF score$\ge$10, we remove the feature with the largest VIF value and recompute VIF values of \emph{n-1} remaining features; \emph{Step 3:} If all VIF values of \emph{n-1} features are less than 10, we stop this process. Otherwise, we repeat step 2.

To remove unimportant features, we used Boruta algorithm \cite{kursa2010feature}. Boruta exploited a Random Forest classification algorithm to measure feature importance score (e.g \emph{z-score}). For each feature, Boruta produces some statistical scores (e.g. mean, max, min, median of \emph{z-score}), and one of three statuses: \emph{confirmed}, \emph{tentative} or \emph{rejected}. \emph{Confirmed} features are important features while \emph{rejected} features are unimportant features. \emph{Tentative} features are those with \emph{insured} importance. We kept only \emph{confirmed} features in our models.

Table~\ref{table:featuresRanking} shows top 10 features at each of the four time points. Project category, goal and the longest reward delivery date were in top 10 features at all the time points. Semantic reward cluster features were in top 10 features at TP1, and also important features at the other times points but not in top 10 features. Temporal features were in top 10 features at TP2 and TP3, whereas creator and backer's activity features were in top 10 features at TP4.

Even though there is no prior work directly related to reward delivery status prediction, we implemented two baselines to compare with our approach:
\squishlist
\item \emph{baseline 1}: It is the majority class selection approach by blindly predicting all projects as late reward delivery projects (54.36\%).






\item \emph{baseline 2:} Kim et al. \cite{kimunderstanding} measured the number of delayed days by 8 features: \# of rewards, goal, project duration, \# of backers, percent raised, \# of backed projects, \# of created projects, and project type. We built XGBoost model with the 8 features as a baseline. Note that \# of backers and percent raised would be available only at TP3 and TP4.
\squishend

To build our predictive model, we built a XGBoost model based on all the features that we listed in Table~\ref{table:features}. We also tried Naive Bayes, SVM and Random Forest based classifiers, but XGBoost classifiers achieved the best results with our features.


\begin{table} [t]
	\centering
	\vspace{-5pt}
	\caption{Prediction results for on-time and late delivery projects. The improvement of XGBoost model over the baselines was significant with p-value $<$ 0.001 using the Wilcoxon directional test.}
\small
	\label{table:AccuracyResult}
	\resizebox{0.36\textwidth}{!}{
	\begin{tabular}{lcccc}
		\toprule
		\multirow{2}{*}{Approach}	& \multicolumn{4}{c}{Accuracy at four time points}        \\
									& TP1 			& TP2 					& TP3 					& TP4 				\\
		\midrule
        Baseline 1					& 54.36\%		& 54.36\%				& 54.36\%				& 54.36\%				\\
		Baseline 2 \cite{kimunderstanding}				& 62.65\%		& 62.65\%				& 63.10\%				& 63.10\%				\\
		
		\midrule
        Our Model           			& \textbf{65.8\%}        & \textbf{66.4\%}    			& \textbf{71.4\%}    	& \textbf{82.5\%}   \\
		\bottomrule
	\end{tabular}
	}
	\vspace{-5pt}
\end{table}

\vspace{-1mm}
\subsection{Experiment Results}
Table~\ref{table:AccuracyResult} shows experimental results of the two baselines and our predictive model. Our XGBoost classifier achieved 65.8\%, 66.4\%, 71.4\% and 82.5\% accuracy at TP1, TP2, TP3, and TP4, respectively. It outperformed the baselines at all the time points, especially, significantly improving the accuracy by 13$\sim$31\% at TP3 and TP4 against the best baseline (p-value $<$ 0.001). The experimental results revealed that achieving high prediction rates at TP1 and TP2 were very difficult with only using limited observable online data available in Kickstarter because of other factors (e.g., factory issues and unexpected problems) even though our model was better than the baselines. However, in the bright side, our model achieved 82.5\% accuracy at TP4, so that it can notify to a project creator, backers and the platform provider whether the rewards delivery will be delayed or not. The creator can announce this news and let backers know in advance (only passing the first 5\% of the longest delivery duration), or the backers can request a refund if they don't want to wait for longer time period.

Next, we analyze which feature group had more distinguishing power between on-time and late delivery projects. In this study, we focus on TP4 containing all the features. In each time, we excluded one of the four feature groups: \emph{creator's activeness}, \emph{backer's activeness}, \emph{linguistic} and \emph{semantic reward cluster} features presented in Table~\ref{table:features}. Then we built a XGBoost classifier based on the remaining features.

Table~\ref{table:FeatureAblationAnalysis} presents experimental results. Removing linguistic features, semantic reward cluster features, backer's activeness features and creator's activeness features reduced the accuracy by 1.7\%, 2\%, 2.4\%, and 10\%, respectively. Overall, creator's activeness features were the most important feature group, even though the other feature groups were also important.

In summary, the experimental results confirmed that our proposed approach was significantly better than the baselines, especially achieving 82.5\% accuracy at TP4.


\begin{table} [t]
	\centering
	\small
	\vspace{-5pt}
	\caption{Feature analysis to understand which feature group degrades our model's performance. The accuracy difference between our model (using all the features) and the rest models are significant with p-value$<$0.01 using two-sample t-test.}
	\label{table:FeatureAblationAnalysis}
	\resizebox{0.36\textwidth}{!}{
	\begin{tabular}{p{2in}c}
		\toprule
		Model 									& Accuracy 	\\
		\midrule
		All											& 82.5\%			\\
		All - \emph{linguistic features}					& 80.8\%			\\
		All - \emph{semantic clustering features} 	& 80.5\%			\\
		All - \emph{backer's activeness}       		& 80.1\%       	\\
		All - \emph{creator's activeness} 			& 72.5\%        	\\
		\bottomrule
	\end{tabular}
	}
	\vspace{-10pt}
\end{table}

\vspace{-1mm}
\section{Predicting Rewards Delivery Duration}
\label{sec:experiment2}


\begin{table*}
	\centering
	\caption{Prediction results estimating delivery duration at four time points. The improvement of our model over the baseline was significant with p-value $<$ 0.001 using the Wilcoxon directional test.}
	\label{table:RegressionResult}
	\resizebox{0.9\textwidth}{!}{
	\begin{tabular}{lrrrr|rrrr|rrrr}
		\toprule
		\multirow{2}{*}{Algorithm}		& \multicolumn{4}{c}{RMSE (days)}        & \multicolumn{4}{c}{NRMSE@A} & \multicolumn{4}{c}{NRMSE@B} \\
							& TP1	& TP2	& TP3	& TP4	& TP1	& TP2	& TP3	& TP4	& TP1	& TP2	& TP3	& TP4		\\
		\midrule
		
		Baseline \cite{kimunderstanding} & 180.0 & 180.0 & 173.0 & 173.0 & 0.248 & 0.248 & 0.239 & 0.239 					& 0.451 & 0.451 & 0.434 & 0.434\\
		\midrule	
		\textbf{Our models}:         & & & & & & & & & & & & \\
		semantic reward cluster features& 106.8 & 106.8 & 106.8 & 106.8 & 0.148 & 0.148 & 0.148 & 0.148  					& 0.268 & 0.268 & 0.268 & 0.268\\
		other features					& 106.1 & 106.1 & 105.6 & 93.2 & 0.146 & 0.146 & 0.145 & 0.128 					& 0.266 & 0.266 & 0.265 & 0.234\\
		All	features & \textbf{100.7}	& \textbf{100.4}	& \textbf{94.1}  & \textbf{78.1}	& \textbf{0.139} & \textbf{0.138}	 & \textbf{0.130} & \textbf{0.108} 	& \textbf{0.252} & \textbf{0.252} & \textbf{0.236} & \textbf{0.196}	\\
		\bottomrule
	\end{tabular}
	}
	\vspace{-5pt}
\end{table*}
\vspace{-1mm}

The previous experimental results motivated us to study how to estimate a project's longest reward delivery duration. What if we can estimate delivery duration accurately at TP1, TP2, TP3 and TP4 in an automated way, it will help the creator to decide better longest reward delivery duration at TP1 or re-estimate it at TP2, TP3 or TP4. In backers' perspective, some backers may be willing to wait longer as long as the project creator notify re-estimated delivery duration in advance (say, TP4) \cite{kimunderstanding}. Therefore, in this section, we build our regression model to estimate delivery duration (in days), and then evaluate its performance compared with 1,598 projects' ground truth described in the Dataset Section.
\subsection{Experimental Setting}
In this study, we used the features presented in Table~\ref{table:features} and applied stepAIC algorithm to choose important features. Then we conducted 10-fold cross-validation for our experiment.

\smallskip
\noindent\textbf{Evaluation Metrics:}
To evaluate each model's performance, we used Root Mean Squared Error (RMSE), and normalized root mean squared error (NRMSE) with regard to the range and the mean of the ground truth data. Lower RMSE and NRMSE values indicate a better model. In the literature, researchers used two versions of NRMSEs, so we also used both of them as evaluation metrics:
\begin{equation}
\resizebox{0.48\textwidth}{!}
{
$
	\text{NRMSE@A} = \frac{RMSE}{max(y) - min(y)}; \text{NRMSE@B} = \frac{RMSE}{\bar{y}}	
$
}
\nonumber
\end{equation}
where $max(y)$, $min(y)$ and $\bar{y}$  are the maximum, the minimum and the mean of the ground truth values, respectively.


\smallskip
\noindent\textbf{Data Transformation:}
Given a project $i$, we denote $h(x^{(i)})$ as the estimated number of days the creator needs to deliver all promised rewards, $y_i$ as the ground truth of the project (e.g. the actual/true number of days that the creator needed to deliver all promised rewards), and $x^{(i)}$ as the feature vector. Before building our regression model, we performed 2-step data transformation as follows:
\begin{itemize}
	\item Transformation on feature values $x^{(i)}$: all feature values in feature vector $x^{(i)}$ of a project $i$ were log transformed. In particular, we used $x^{(i)} = log(1 + x^{(i)})$ instead of using the original feature values.

	\item Transformation on ground truth $y_i$: we used box-cox transformation \cite{kutner2004applied} to transform $y_i$ as follows:
	\[
		y^{(i)}_{new} = \begin{cases}
								\frac{y_i^\lambda - 1}{\lambda}, &\text{if $\lambda \neq 0$} \\
								log(y_i), 						 &\text{if $\lambda = 0$}
						\end{cases}
	\]
\end{itemize}

To choose the best value of $\lambda$, we fit a multiple linear regression $h(x) = \sum_{i=0}^{n}(\theta_ix_i + \epsilon_i) = \theta^Tx + \epsilon$. 
We assumed all the errors $\epsilon$ are independent and $\epsilon \sim \mathcal{N}(0,\sigma^2)$. After doing transformation, the log-likelihood $\mathcal{L}$ of the model, with regard to the value of $\lambda$, is calculated as below:
 \[
	 \mathcal{L} = -\frac{n}{2}log \bigg[ \sum_{i=1}^{m}\bigg(\frac{y_{new} - h_\theta(x^{(i)})}{exp (\frac{1}{n}\sum_{i=1}^{m} log(y_i)) ^\lambda } \bigg) ^ 2 \bigg]
 \]

We varied the value of $\lambda$ in the grid [-1, 1] and selected $\lambda=0.11$ since it maximized the log-likelihood $\mathcal{L}$. However, if we transform $y_{new} = y^\lambda$, the magnitude of the sum of squared error (SSE) will be changed. In other words, the value of SSE will depend on $\lambda$. To overcome this issue, we normalized $y_{new}$ by the geometric mean of all raw values $y_i$ as below:
\[
\vspace{-3pt}
y_{new} = \frac{(y_i^\lambda - 1)}{\lambda \prod_{i=1}^{n}{(\sqrt[n]{y_i})} ^{\lambda-1}}
\vspace{-3pt}
\]


\smallskip
\noindent\textbf{Our Model:}
To predict how many days the creator of a project $i$ needs to fully deliver all the promised rewards, we built a multiple linear regression model based on the features in Table~\ref{table:features} except the \emph{longest reward delivery duration} feature. Let $x$ be the feature vector ($x_0 = 1$), $m$ be $|$projects$|$, $\theta$ be the coefficient vector of the feature vector (except that $\theta_0$ is the intercept), and $\epsilon_i$ be the error term which follows normal distribution. We used the squared loss and introduced elastic net regularization to find the optimal values of the coefficient vector $\theta$ by minimizing the following loss function:
\begin{equation}
\vspace{-3pt}
\label{equation:CostFunction}
min\underset{\theta}{\mathcal{L}} = \frac{1}{2}\sum_{i = 1}^{m} (\sum_{j=1}^{n}\theta_{ij}x_{ij} - y_i)^2 + \lambda_1||\theta||_1 + \lambda_2||\theta||^2_2
\vspace{-3pt}
\end{equation}
Here, $\lambda_1$ and $\lambda_2$ were used to control the regularization effect.

\smallskip
\noindent\textbf{Baseline:}
Kim et al. \cite{kimunderstanding} proposed 8 features and built a simple multiple linear regression to measure the number of delayed days (e.g. the difference in days between estimated reward delivery date and real reward delivery date). 
\vspace{-1mm}
\subsection{Experiment Results}
Table~\ref{table:RegressionResult} shows experimental results of all the methods at TP1, TP2, TP3, and TP4. We implemented 3 models with different feature sets: (i) \emph{semantic reward cluster features}; (ii) \emph{other features}: all features presented in Table \ref{table:features} except semantic reward cluster features and \emph{longest reward delivery duration} feature; and (iii) \emph{all features}: combining (i) and (ii). All of our models outperformed the baseline. Interestingly, the \emph{semantic reward cluster features} model achieved almost same result with the \emph{other features} model at TP1, TP2 and TP3. It indicates that the semantic reward cluster features were helpful in estimating the reward delivery duration. Adding available features at TP4 (e.g. creator's activeness features + backer's activeness features + linguistic features) reduced 12\% error of the \emph{other features} model. The \emph{all features} model gained the best result with lowest RMSE=100.7, NRMSE@A=0.139, and NRMSE@B=0.252 at TP1, significantly reduced 44\% error compare to the baseline. It also significantly reduced RMSE, NRMSE@A and NRMSE@B by minimum 45\% and 55\% at TP3 and TP4, respectively compared with the baseline. Comparing with NRMSE@A of another model in another domain \cite{shahamiri2014real}, our model performed well, indicating the effectiveness of our regression model. We also tried SVM, Neural Network and Random Forest based on our features to build regression models. But, the multiple linear regression with the elastic net regularization got the best result.

In this context, RMSE=78.1 means the average difference between our estimated delivery duration and the ground truth (i.e., actual delivery duration) is 78.1 days. In a real-world scenario, project creators at TP1 can use our model to estimate delivery duration. Then, they can add 101 days to the estimated delivery duration as a buffer to make sure they can get enough delivery duration. Similarly, creators at TP4 can get estimated delivery duration from our model, and add 78 days as a buffer to the estimated delivery duration. The final delivery duration will be safe/enough delivery duration to deliver reward within the longest delivery duration.

We further investigated to understand whether there is any significant difference between the most correctly predicted projects and the most incorrectly predicted projects. In particular, we extracted top 10 correctly predicted projects and top 10 incorrectly predicted projects. Then we conducted Wilcoxon test which showed that top 10 correctly predicted projects provided a larger number of projects and contained a larger number of sentences in their reward descriptions compared with top 10 incorrectly predicted projects.

\vspace{-1mm}
\section{Conclusion}


In this paper, we analyzed Kickstarter projects, and behavior of creators and backers to find distinguishing patterns between on-time and late reward delivery projects. Then, we proposed a clustering approach to group rewards by their latent difficulty levels. Based on the analysis and study, we extracted various features toward building predictive models for a reward delivery status (i.e., on-time or late) and delivery duration. Our experimental results show that our predictive model outperformed two baselines, achieving 71.4\% and 82.5\% accuracy at TP3 and TP4, respectively and improving the accuracy by 13$\sim$31\%. In real-world use cases, the predictive model will help project creators to know whether their rewards will be delivered most likely on-time or late. They can announce this news in advance, only passing the first 5\% of the longest reward delivery duration with 82.5\% accuracy. Backers can decide whether they are going to wait further or request a refund at that point without worrying about the worst case (e.g., hearing the delay 1 year later).

Our regression model to estimate delivery duration also outperformed the baseline, achieving 0.130 and 0.108 NRMSE@A at TP3 and TP4, respectively, and reducing NRMSE@A by 45$\sim$55\%. Project creators can make better delivery duration estimation by using our regression model. Backers can also know when rewards will be most likely delivered. Platform providers can provide these predictive and regression models as a tool to help creators and backers. Therefore, our research will improve the trustworthiness among creators, backers and platform providers. Since it is the first work to predict reward delivery status and delivery duration in reward-based crowdfunding platforms, it opens the door of a new research problem to the research community, and provides our models as new baselines. In addition, researchers may come up with other research problems in the \emph{reward delivery phase}. 



\section*{Acknowledgment}
This work was supported in part by NSF grant CNS-1553035. Any opinions, findings and conclusions or recommendations expressed in this material are the author(s) and do not necessarily reflect those of the sponsors.

{\small	
\bibliographystyle{IEEEtran}
\bibliography{IEEEabrv}
}

\end{document}